\documentclass[aps,prd,showpacs,showkeys,superscriptaddress,twocolumn]{revtex4-1}
\usepackage[colorlinks,linkcolor=blue,anchorcolor=blue,citecolor=blue,urlcolor=blue,breaklinks=true]{hyperref}
\usepackage{graphicx}% Include figure files
\usepackage{amsmath}
\usepackage{amssymb}
\usepackage{slashed}
\usepackage{latexsym}
\usepackage{epsfig}
\usepackage{amsbsy}
\usepackage{array}
\usepackage{amssymb}
\usepackage{setspace}
\usepackage{bm}
\usepackage{lipsum}
\usepackage{mathrsfs}
\usepackage{float}
\usepackage{color}
\usepackage{booktabs}
\usepackage[T1]{fontenc}
\usepackage{mathptmx}
\usepackage{changes}
\DeclareMathAlphabet{\mathcal}{OMS}{cmsy}{m}{n}
\DeclareSymbolFont{largesymbols}{OMX}{cmex}{m}{n}

\begin{document}
\author{Li-Qun Su}
\email{xzslq1203@smail.nju.edu.cn}
\affiliation{Department of physics, Nanjing University, Nanjing 210093, China}
\author{Chao Shi}
\email{cshi@nuaa.edu.cn}
\affiliation{Department of Nuclear Science and Technology, Nanjing University of Aeronautics and Astronautics, Nanjing 210016, China}
\author{Yong-Feng Huang}
\email{hyf@nju.edu.cn}
\affiliation{School of Astronomy and space science, Nanjing University, Nanjing 210023, China}
\author{Yan Yan}
\email{2919ywhhxh@163.com}
\affiliation{School of Mathematics and physics, Changzhou University,Changzhou 213164, China}
\author{Cheng-Ming Li}
\email{licm.phys@gmail.com}
\affiliation{School of physics and Microelectronics, Zhengzhou University, Zhengzhou 450001, China}

\author{Hongshi Zong}
\email{zonghs@nju.edu.cn}
\affiliation{Department of physics, Nanjing University, Nanjing 210093, China}
\affiliation{Department of physics, Anhui Normal University, Wuhu 241000, China}
\affiliation{Nanjing Proton Source Research and Design Center, Nanjing 210093, China}
\affiliation{Joint Center for Particle, Nuclear Physics and Cosmology, Nanjing 210093, China}
\date{\today}
	
\title{Quark stars in the pure pseudo-Wigner phase}
	
\begin{abstract}
In this paper, the equation of state (EOS) of deconfined quark stars is studied in the framework of the two-flavor NJL model, and the self-consistent mean field approximation is employed by introducing a parameter $\alpha$ combining the original Lagrangian and the Fierz-transformed Lagrangian, $\mathcal{L}_R= (1-\alpha)\mathcal{L}+\alpha\mathcal{L}_F$, to measure the weights of different interaction channels. It is believed that the deconfinement of phase transition happens along with the chiral phase transition. Thus, due to the lack of description of confinement in the NJL model, the vacuum pressure is set to confine quarks at low densities, which is the pressure corresponding to the critical point of chiral phase transition. We find that deconfined quark stars can reach over two-solar-mass, and the bag constant therefore shifts from $(130 ~\mathrm{MeV})^4$ to $(150 ~\mathrm{MeV})^4$ as  $\alpha$ grows.
In addition, the tidal deformability $\Lambda$ is yielded ranging from 253 to 482 along with the decrease of $~\alpha$, which satisfies the astronomical constraint of $\Lambda<800$ for 1.4-solar-mass neutron stars.
\bigskip

\end{abstract}
	
\maketitle

\section{Introduction}

Neutron stars are some of the densest manifestations of massive objects in the universe. They are natural and ideal astrophysical laboratories for testing theories of strongly interacting matter.
With the help of pulsar observations, the maximum mass of observed neutron stars is inceasing, i.e., PSR J0348 + 0432 \cite{Antoniadis1233232} with 2.01$\pm$0.04 $M_{\odot}$, and therefore the equation of state (EOS) is constrained since soft EOS can not support massive stars. Besides, the measurement of radius of neutron stars is obtained with x-ray observations \cite{Bogdanov_2019,Riley_2019,Capano_2020}, such as $11.0^{+0.9}_{-0.6}~$km for 1.4-solar-mass compact stars in Ref. \cite{Capano_2020}. After the detection of the Gravitational Wave, i.e., the GW170817 of Binary Neutron Star (BNS) merging event \cite{PhysRevLett.119.161101}, the tidal deformability provides an extra constraint on the EOS, $\Lambda<800 $ for the 1.4-solar-mass compact stars, which rules out many stiff EOS with large tidal deformabilities. Considering these strict constraints, it is commonly believed that
neutron stars encompass ``normal'' stars \cite{Wu:2013hqa,PhysRevD.86.114028,Li:2017xlb,Li:2018ltg,Li:2018ayl} and ``strange quark matter'' stars \cite{WittenPRD1984}. The concept of strange quark matter stars originated from Witten's strange quark matter hypothesis. However, so far, Witten's strange quark matter hypothesis has neither been proven nor falsified.
%the most possible structure of the compact stars consists of hadronic matter in the outside layer and deconfined quark matter in the core \cite{Wu:2013hqa,PhysRevD.86.114028,Li:2017xlb,Li:2018ltg,Li:2018ayl}, namely hybrid star. However, these constraints allow the possibilities of deconfined quark stars. Because the stability between u-d quark matter and u-d-s quark matter is still an open question, the two-flavor quark star and three-flavor quark star are both the candidates for compact stars. It was first proposed by witten \cite{WittenPRD1984} that the energy per quark in the strange-quark matter (SQM) is less than the energy per quark of the most stable atomic nucleus, ${}^{56}\mathrm{Fe}$. Therefore, Many authors show that SQM can be the stable state to form the quark star \cite{PhysRevD.101.063023,PhysRevD.43.627,PhysRevC.62.025801,PhysRevD.92.084009,PhysRevD.99.043001}.

Recently, in Ref.  \cite{ChenZhangPRL2018}, after taking into account of the bulk effect, baryonic matter with only u and d quarks can also be the stable ground state for the baryon number $A > 300$. Thus, the non-strange quark matter can be the ground state of baryonic matter rather than the strange quark matter \cite{zhaotong,PhysRevD.100.123003,doi:10.1143/JPSJ.58.3555,doi:10.1143/JPSJ.58.4388,PhysRevD.4.1601,PhysRevD.101.043003}.
Therefore, it is extremely interesting to study whether there are neutron stars composed of deconfined non-strange quark matter in nature. The main purpose of this paper is to discuss the possibility of such neutron stars.
%Therefore, the deconfined quark star with u-d quark matter is not excluded by the astronomic observation, and may exist in the universe.

In this paper, the two-flavor Nambu--Jona-Lasinio(NJL) model \cite{PhysRevD.9.3471,Alford:2004pf,Roberts:1994dr,Roberts:2000aa,Maris:2003vk,Cloet:2013jya,PhysRevD.90.114031,Wang:2015tia,Xu:2015vna,Asakawa:1989bq,Klevansky:1992qe} will be used to describe the quark matter in neutron stars, and it is established to manifest the spontaneous breaking of chiral symmetry, where the effective quark mass is acquired at low densities.
As the chemical potential increases, the chiral phase transition occurs and the chiral symmetry begins to restore, which is associated with the fact that the hadronic matter transits to the quark matter, corresponding to the Nambu phase and the pseudo-Wigner phase respectively \cite{Xu_2018,PhysRevD.99.076006}.
The pressure  of nuclear matter then reads \cite{PhysRevD.78.054001,Zong:2008zzb},
\begin{align}
P(\mu)=P(\mu=0)+\int_0^{\mu} \rho(\mu') d\mu', \nonumber
\end{align}
where $\rho(\mu')$ represents the particle number density at chemical potential $\mu'$.
The vacuum pressure $P(\mu=0)$ is set as $-B$ (bag constant) to confine quark matter at zero chemical potential. The EOS of strong interaction matter can then be obtained.
In our previous works of Refs. \cite{PhysRevD.101.063023,zhaotong,PhysRevD.100.123003}, an improved
NJL model is engaged to derive the EOS of the neutron star, which ranges from Nambu phase to pseudo-Wigner phase in our calculation.  The calculated neutron stars have nucleonic matter on its surface, which transforms into deconfined quark matter in their cores. So they are indeed hybrid stars rather than pure quark stars. Following the hadron-quark duality, people can in principle use equivalent quark degree of freedom to describe the properties of strongly interacting matter in low-density regions. This means that in principle we can uniformly describe the strong interaction matter EOS from low density to high density with equivalent quark degree of freedom. In this work, we discuss the possibility of neutron stars being pure quark stars, i.e., with quark matter of pseudo-Wigner phase on its surface, hence the hadronic matter is totally absent.

Additionally, in lattice simulations, a single transition is observed at temperature. It indicates that the deconfinement transition and the chiral transition may happen at the same time \cite{PhysRevLett.110.172001}. Hence, the confinement and chiral phase transition are normally considered along with each other at finite densities as well, even though there is no strong evidence for such relations. For these reasons, quark stars should consist of quark matter with only the pseudo-Wigner phase, and the EOS is contributed only with the chemical potential over the critical point $\mu_c$ of the chiral phase transition. We should note that the pure deconfined quark matter has tremendous pressure which is required to suppress for stability. Fortunately, the non-trivial QCD vacuum is able to support the huge pressure, just like quarks confined by the vacuum as hadrons, and, in conclusion, the rest parts of the pressure which are contributed from the chemical potential below $\mu_c$ is redefined as the vacuum pressure. In this case, quark matter becomes a self-constrained system, and the astronomical observations do not eliminate the possibilities of deconfined quark stars.

In this paper, a self-consistent NJL model is employed, which is recently proposed in Refs. \cite{Yu:2020dnj,PhysRevD.100.094012,PhysRevD.100.043018,PhysRevD.92.084009,PhysRevD.100.123003,Wang_2019,PhysRevD.100.123003,PhysRevD.100.123003}. In this model, the parameter $\alpha$ is introduced to evaluate the competition between the vector interaction channel and the scalar interaction channel \cite{altland2010condensed}. With the increase of $\alpha$, corresponding to the stronger repulsive interactions between quarks, quark matter can support large pressures, and therefore a massive quark star over two-solar mass becomes possible. In addition, the proper-time regularization is adopted for the reason that the three-momentum cutoff has disadvantages in dealing with the quark matter at large densities, since the central chemical potential for massive quark stars is usually beyond the cutoff.

This paper is organized as follows:
In Section \ref{section:2},
the self-consistent NJL model is adopted and the gap equation with the parameter $\alpha$ is given.
In Section \ref{section:3},
the effective quark mass and particle number density are presented  with the proper-time regularization.
In Section \ref{section:4},
the EOS in the framework of the NJL model is calculated under the condition of electric charge neutrality.
In Section \ref{section:5},
the TOV equation is solved and the mass-radius relations of quark stars with different parameter $\alpha$ are shown.
In Section \ref{section:6},
the tidal deformability of quark stars is calculated which is found to satisfy
the constraints from astronomical observations.
In Section \ref{section:7},
our conclusions are presented.

\section{THE SELF-CONSISTENT NJL MODEL}\label{section:2}

The standard Lagrangian of two-flavor NJL model with chemical potential is,
\begin{align}
\mathcal{L}=\bar{\psi}(i\gamma^\mu\partial_\mu-m_0+\mu\gamma^0)\psi+G[(\bar{\psi}\psi)^2+(\bar{\psi}i\gamma_5\boldsymbol\tau\psi)^2],\label{eq:1}
\end{align}
where $G$ is the four-fermion coupling constant, $m_0$ is the current quark mass
matrix and$~\boldsymbol\tau~$is the Pauli matrix.
The mean field approximation is employed \cite{Buballa:2003qv},
\begin{align}
\mathcal{L}_\mathrm{eff}=\bar{\psi}(i\gamma^\mu\partial_\mu-m_0+2G\sigma+\mu\gamma^0)\psi-G\sigma^2,\label{eq:2}
\end{align}
here $~\sigma=\langle\bar{\psi}\psi\rangle~$ is the quark condensate. The effective quark mass is obtained as
\begin{align}
M=m_0-2G\sigma.\label{eq:3}
\end{align}
The quark condensate is derived as
\begin{align}
\sigma=\langle\bar{\psi}\psi\rangle=\int\dfrac{d^4p}{(2\pi)^4}\mathrm{Tr}[S(p)],\label{eq:4}
\end{align}
here $S(p)$ represents the dressed quark propagator, and the trace operates on the Dirac, flavor and color space.
Under the circumstances of zero temperature and finite chemical potential,
the quark condensate reduces to:
\begin{align}
\sigma=-2N_cN_f\int \dfrac{d^3p}{(2\pi)^3}\dfrac{M}{E_p}\big[1-\theta(\mu-E_p)\big].\label{eq:6}
\end{align}
where $E_p=\sqrt{p^2+M^2}$.

For four-fermion intraction channels, the direct and exchange interactions are associated with the Fierz-transformation, and the Fierz-transformation in the framework of the NJL model is
\begin{align}
\mathscr{F}=&\dfrac{1}{8\mathrm{N_c}}[2(\bar{\psi}\psi)^2+2(\bar{\psi}i\gamma_5\boldsymbol\tau\psi)^2-2(\bar{\psi}\boldsymbol\tau\psi)^2-2(\bar{\psi}i\gamma_5\psi)^2\nonumber\\
&-4(\bar{\psi}\gamma^{\mu}\psi)^2-4(\bar{\psi}i\gamma^{\mu}\gamma_5\psi)^2+(\bar{\psi}\sigma^{\mu\nu}\psi)^2-(\bar\psi\sigma^{\mu\nu}\boldsymbol\tau\psi)^2],\label{eq:7}
\end{align}
where color terms are neglected. We only keep the scalar and the vector interaction channels, which is most crucial for the EOS of neutron stars.
Because the original Lagrangian and Fierz-transformed Lagrangian are identical, they can be combined at any proportions. The weighting factor $\alpha$ reflects the competition between the scalar interaction channels and the vector interaction channels,
\begin{align}
\mathcal{L}_R= (1-\alpha)\mathcal{L}+\alpha\mathcal{L}_F,\label{eq:8}
\end{align}
here $\mathcal{L}$ is the original Lagrangian and $\mathcal{L}_F$ is the Fierz-transformed Lagrangian. However, this situation is different when we employ the mean field approximation, because the Fierz-transformation and mean-field approximation are not commutative:
\begin{align}
&M=m_0-2G(1-\alpha+\dfrac{\alpha}{4N_c})\sigma,\label{eq:9}\\
&\mu'=\mu-\dfrac{G'\alpha}{N_c\pi^2}\langle\psi^\dagger\psi\rangle, \label{eq:10}
\end{align}
where $~G'=\dfrac{2G(1+\dfrac{1}{4N_c})}{(1-\alpha+\dfrac{\alpha}{4N_c})}~$.

It is clear that the gap equations are affected by $\alpha$ \cite{Wang2019}, and the vector interaction between quarks becomes dominating with the increase of $\alpha$.  The ``correct'' choice of ratio can be motivated only by physical reasoning, not by plain mathematics \cite{altland2010condensed}. We will show that the parameter $\alpha$ is crucial for the EOS of quark stars, and can be constrained by astronomical observations.

\section{PROPER-TIME REGULARIZATION}\label{section:3}

As indicated by the divergence of quark condensates, the three-momentum cutoff is usually adopted. If the quark matter can support at least two-solar mass, the densities inside the quark star are extremely large and therefore the central chemical potential of quark stars may reach over 700 MeV.
In this region, the $\theta$ function of the equation is malfunctioning,
because the chemical potential is beyond the cutoff of the momentum.
We have the mathematical identity of
\begin{align}
\dfrac{1}{A^n}=\dfrac{1}{\Gamma(n)}\int_0^{\infty}d\tau \tau^{n-1}e^{-\tau A} \rightarrow \dfrac{1}{\Gamma(n)}\int_{\tau_{UV}}^{\infty}d\tau \tau^{n-1}e^{-\tau A}, \label{eq:11}
\end{align}
where $\tau_{UV}$ is set to regularize the divergence. Thus the proper-time regularization is employed and the quark condensate now can be altered as
\begin{align}
\sigma&=-2N_cN_f\Big[\int \dfrac{d^3p}{(2\pi)^3}M\int_{\tau_{UV}}^{\infty}d\tau \dfrac{1}{\sqrt{\pi\tau}}e^{-\tau E_p^2} \nonumber\\
&~~~~~-\int \dfrac{d^3p}{(2\pi)^3}\dfrac{M}{E_p}\theta(\mu-E_p)\Big]\nonumber\\
&=-2N_cN_f\Big[\int_{\tau_{UV}}^{\infty}d\tau \dfrac{M}{\tau^2}e^{-\tau E_p^2}-\int \dfrac{d^3p}{(2\pi)^3}\dfrac{M}{E_p}\theta(\mu-E_p)\Big].\label{eq:12}
\end{align}

\begin{figure}[t]
	\centering
	\includegraphics[width=1\linewidth]{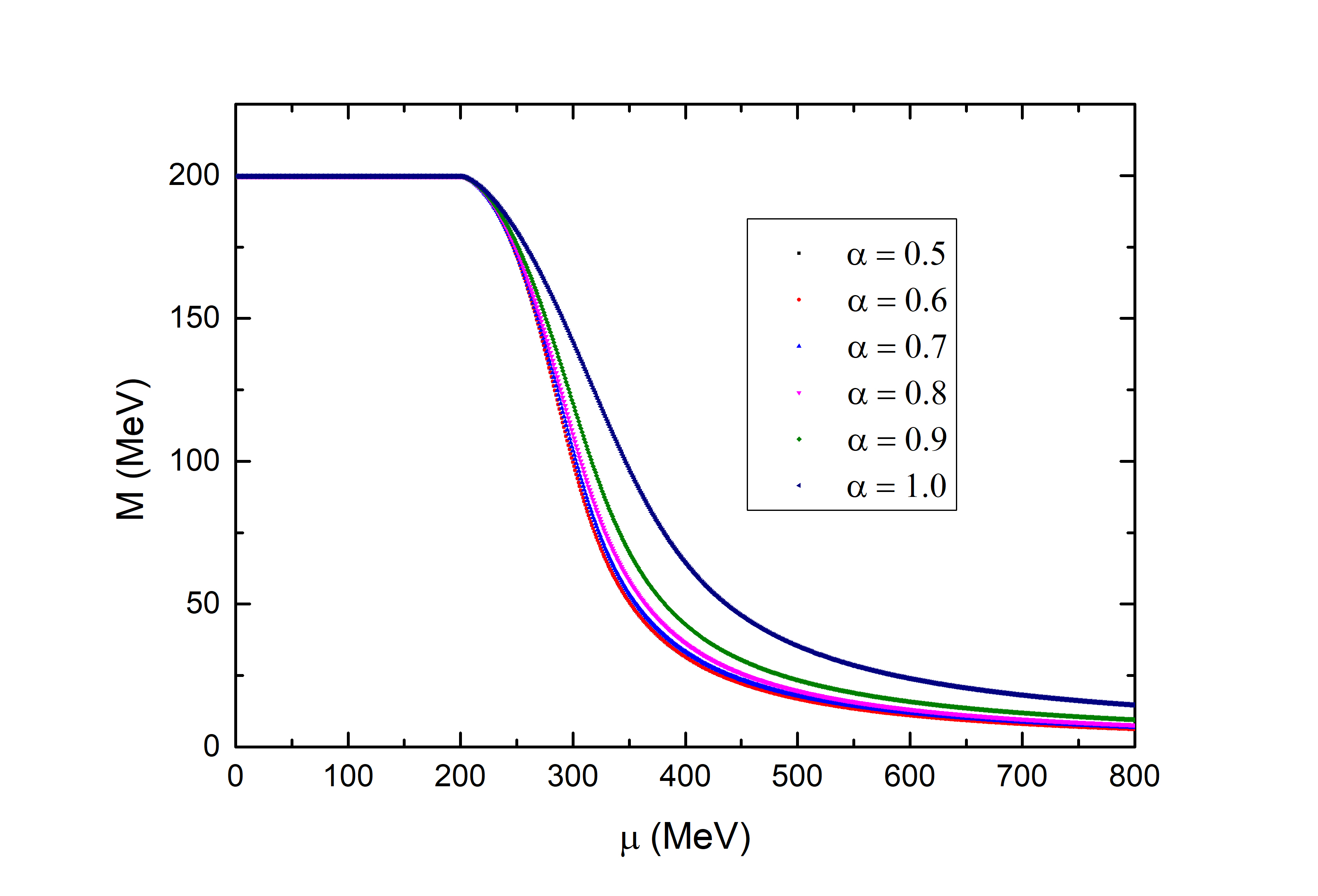}\\
	\caption{The effective quark mass as a function of the chemical potential.}\label{fig:mass}
\end{figure}

In this study, the parameters will be fixed as $\mathrm{m_0}=5.0 ~\mathrm{MeV}, ~\mathrm{\tau_{UV}}=1092^{-2}~\mathrm{MeV}$ and ~$\mathrm{G'}=3.086 \times 10^{-6} ~\mathrm{MeV}^{-2}$,
which fit the pion decay constant and the pion mass at zero temperature and chemical potential.
The solutions of the gap equations are presented in Fig. \ref{fig:mass}.
The effective quark mass is 199.73 MeV at zero chemical potential, which is resulted from the dynamical breaking of chiral symmetry and it describes the hadronic matter in quark degrees of freedom, corresponding to the Nambu phase. As a result, the quark condensate associated with the breaking of chiral symmetry reflects strong interactions between quarks in the non-trivial QCD vacuum.
With the increase of chemical potentials, the effective quark mass declines and the chiral symmetry starts to restore. When the chemical potential of quark matter is large enough, the effective quark mass shifts to the current quark mass due to asymptotic freedom, corresponding to the pseudo-Wigner phase.

\begin{figure}[H]
	\centering
	\includegraphics[width=1\linewidth]{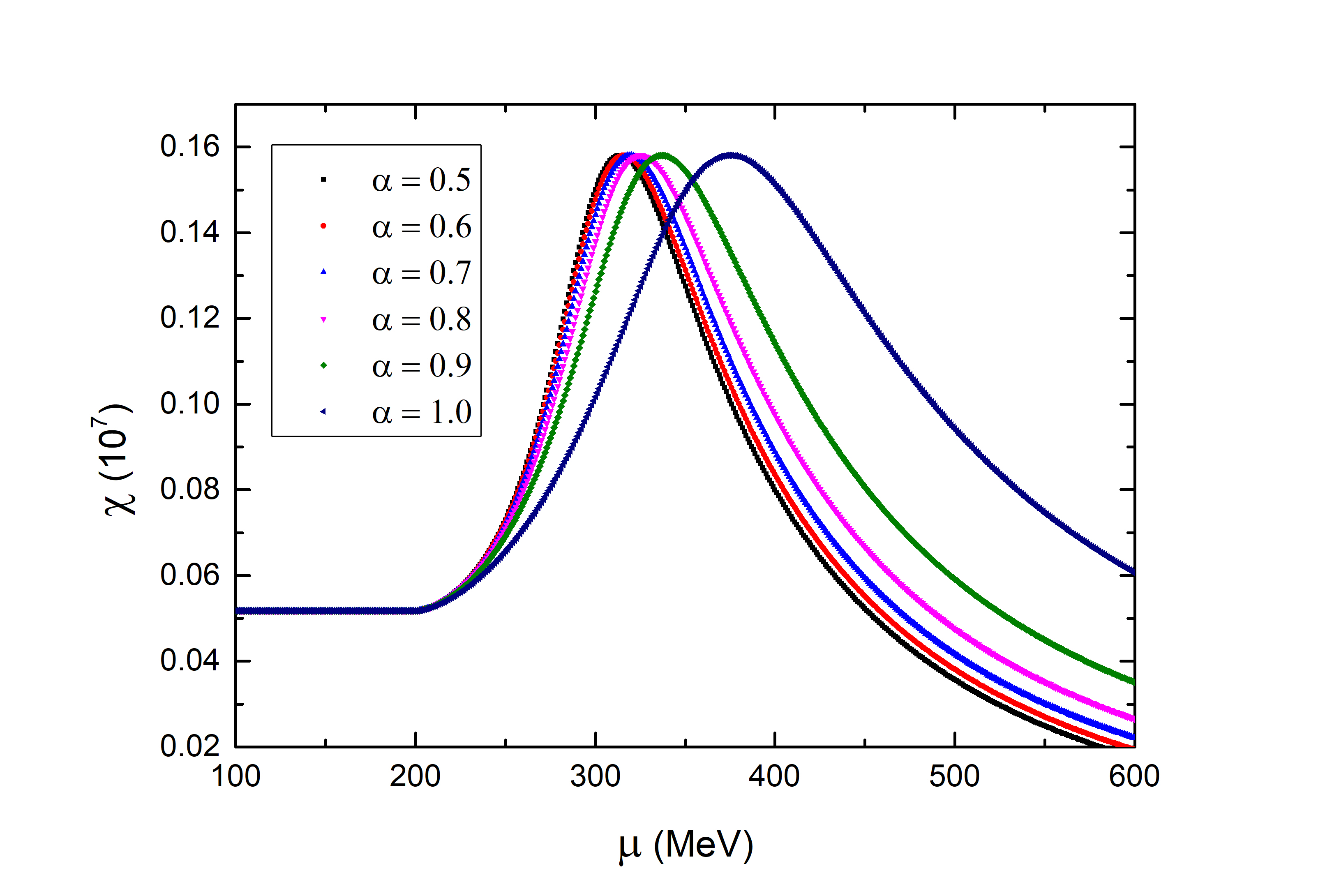}\\
	\caption{The chiral susceptibility as a function of the chemical potential for different $\alpha$.}\label{fig:sus}
\end{figure}

The chiral susceptibility is calculated by \cite{PhysRevD.58.096007}
\begin{align}
\chi=\dfrac{\partial\sigma}{\partial m_0}.\label{eq:13}
\end{align}
As is shown in Fig. \ref{fig:sus}, the finite and smooth peak of the chiral susceptibility manifests
that the phase transition is the crossover. Besides, the critical value of the chemical potential is
$\mu_c=313,~316,~319,~325,~337,~375 ~\mathrm{MeV}$ for $\alpha= 0.5,~0.6,~0.7,~0.8,~0.9,~1.0$, respectively.

\begin{figure}[t]
	\centering
	\includegraphics[width=1\linewidth]{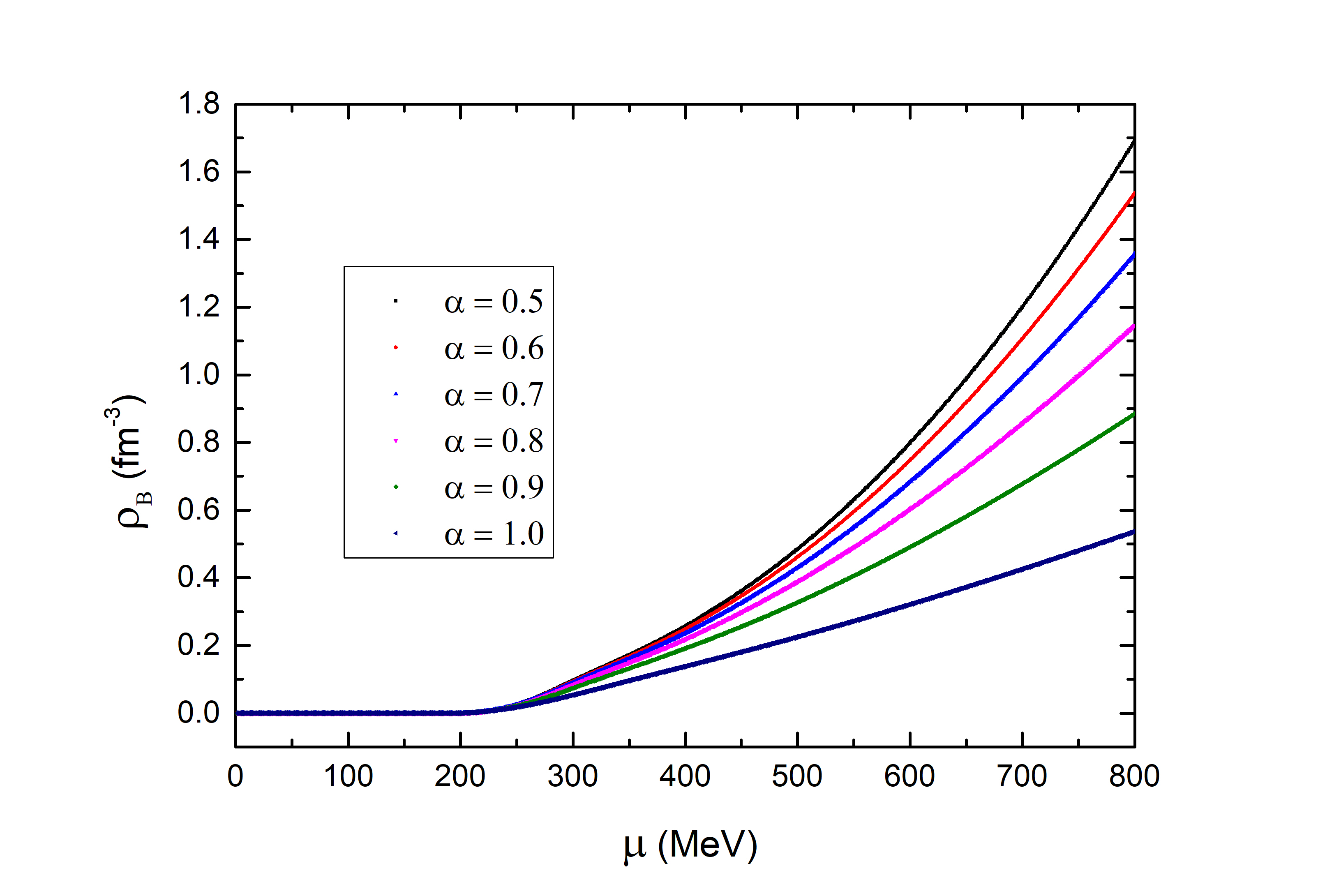}\\
	\caption{The baryon number density as a function of the chemical potential for different $\alpha$.}\label{fig:rho}
\end{figure}

Furthermore, the quark number density at zero temperature is derived as
\begin{align}
\langle\psi^\dagger\psi\rangle=2N_cN_f\int \dfrac{d^3p}{(2\pi)^3}\theta(\mu-E_p).\label{eq:14}
\end{align}
The baryon number densities for different $\alpha$ are shown in Fig. \ref{fig:rho}.
It entails that the quark matter appears at  $~\mu_0=200 ~\mathrm{MeV}$.
At a fixed chemical potential, we can see that the baryon number density decreases as $\alpha$ increases.
When the chemical potential reaches 800 MeV,
the baryon number density is~$ 1.69,~1,53,~1.35, ~1.15,~ 0.89, ~0.54 ~\mathrm{fm}^{-3}$
for $\alpha= 0.5,~ 0.6, ~0.7, ~0.8,~ 0.9, ~1.0$, respectively.

\section{THE EQUATION OF STATE}\label{section:4}

\begin{figure}[H]
	\centering
	\includegraphics[width=1\linewidth]{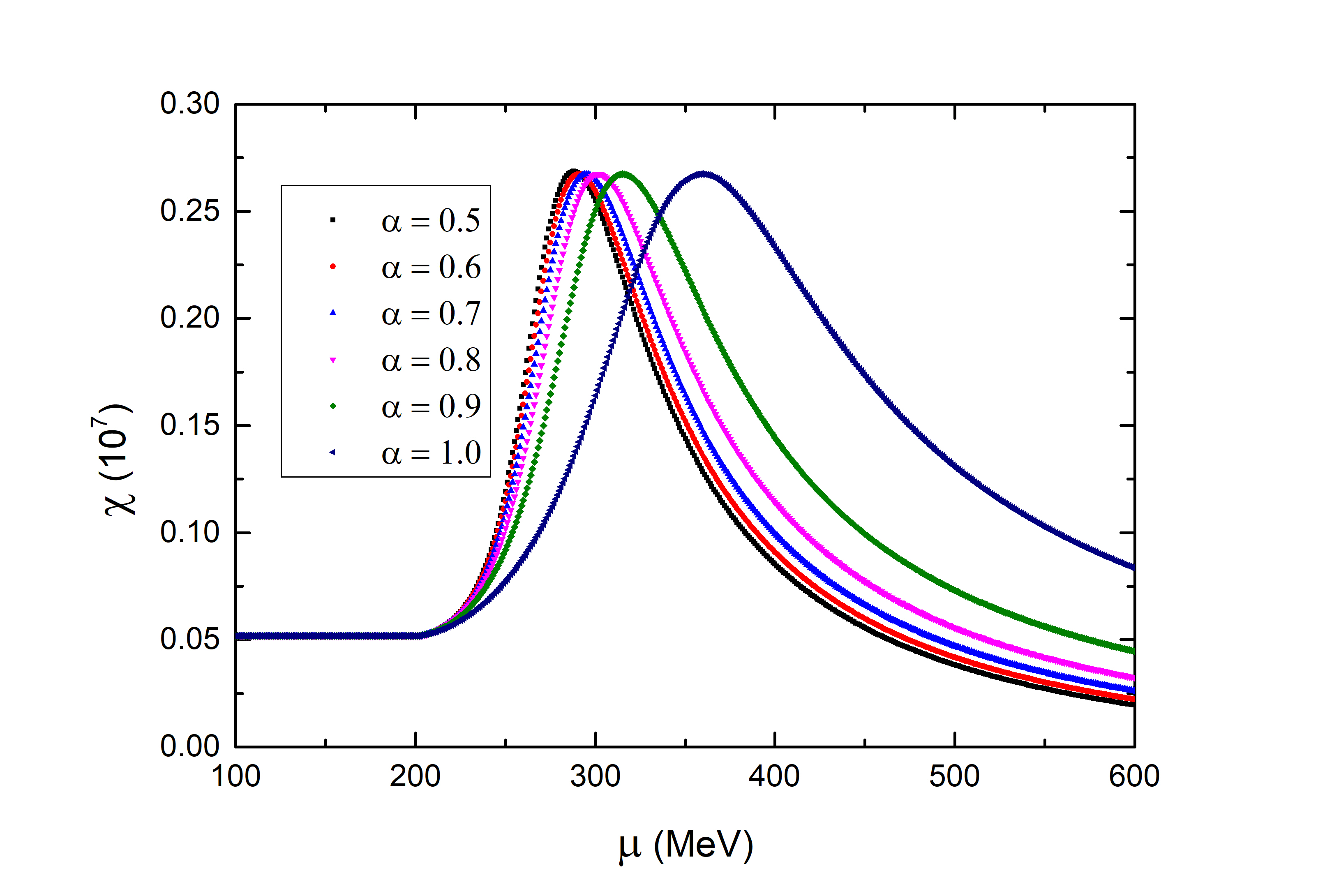}\\
	\caption{The chiral susceptibility as a function of the chemical potential for different $\alpha$, when electric and chemical equilibrium conditions are satisfied.}\label{fig:susN}
\end{figure}

The self-consistent NJL model with the proper-time regularization has been presented. However, a real quark star has to be in $\beta$ equilibrium, and the electric charge neutrality should be satisfied. Thus, the boundary conditions are included as
\begin{align}
\mu_d&=\mu_u+\mu_e,\label{eq:15}\\
\dfrac{2}{3}\rho_u&=\dfrac{1}{3}\rho_d+\rho_e,\label{eq:16}
\end{align}
where $\rho_u,~\rho_d, ~\rho_e$ are the quark number densities of u-quarks, d-quarks and electrons.
Under these conditions, the chiral susceptibility is re-calculated and the results is
shown in Fig. \ref{fig:susN}. Comparing with Fig. \ref{fig:sus},
the chiral transitions are still crossover, and the peaks are
at $\mu_c=288,~291,~ 295, ~302, ~316, ~359 ~\mathrm{MeV}~$
for $\alpha= 0.5, ~0.6,~ 0.7, ~0.8, ~0.9,~ 1.0$, respectively.
We see that the critical points of chiral phase transition become smaller
when the charge neutrality is taken into account.

For the reason of the asymmetry between u-quark and d-quark in the beta equilibrium system,
the pressure of quark matter is given as \cite{PhysRevD.78.054001,Zong:2008zzb}:
\begin{align}
P(\mu_u, \mu_d)=&P(\mu_u=0,\mu_d=0)+\int_0^{\mu_u} \rho_u(\tilde{\mu}_u,\mu_d=0)d\tilde{\mu}_u\nonumber\\
&+\int_0^{\mu_d} \rho_d(\mu_u,\tilde{\mu}_d)d\tilde{\mu}_d.\label{eq:17}
\end{align}
The first term on the right-hand side is the contribution of the quark matter at zero temperature and zero chemical potential,
corresponding to the vacuum pressure. In the case of the MIT bag model,
the pressure of the quark matter is defined as:
\begin{align}
P(\mu)=P_{\mathrm{free}}(\mu)-B,\label{eq:18}
\end{align}
where $\mathrm{P_{free}}(\mu)$ refers to the pressure contributed from free quark matter.
In this case, the bag constant $B$ is set to keep free quarks constrained inside the bag, which is due to the fact that free quarks are never observed. Thus the quark matter should have negative pressure at zero chemical potential.

When the pressure of quark matter in the framework of the MIT bag model is beyond zero
along with the increase of chemical potentials,
free quarks are no longer trapped inside the bag.
Thus, the bag constant is yielded as the pressure corresponding to the critical point of deconfinement,
\begin{align}
B=P_{\mathrm{free}}(\mu_{de})=\int_0^{\mu_{de}}\rho_{\mathrm{free}}(\tilde{\mu})d\tilde{\mu},\label{eq:19}
\end{align}
where $\mu_{de}$ represents the chemical potential under which free quark matter appears.
In the case of the NJL model, the bag constant should have the same physical meaning.
It is assumed that the chiral phase transition and deconfinement transition happen roughly at the same time \cite{PhysRevLett.110.172001}. As a result, we redefine the bag constant as the pressure when chiral phase transition begins so as to make sure that quark matter on the surface starts to appear with zero pressure. Then we have
\begin{align}
&P(\mu_c)=P(\mu=0) + \int_0^{\mu_c}\rho d \tilde{\mu}=0,\nonumber\\
&P(\mu=0)= - B =-\int_0^{\mu_c}\rho d \tilde{\mu}.\label{eq:20}
\end{align}
In conclusion, the pressure of the quark matter in the framework of the two-flavor NJL model is derived as:
\begin{align}
P(\mu_u, \mu_d)=&-B+\int_0^{\mu_u} \rho_u(\tilde{\mu}_u,\mu_d=0)d\tilde{\mu}_u\nonumber\\
&+\int_0^{\mu_d} \rho_d(\mu_u,\tilde{\mu}_d)d\tilde{\mu}_d,\label{eq:21}
\end{align}
where$
~B=\sum_{f}\int_0^{\tilde{\mu}_c}\rho_f d \tilde{\mu}_f$.

In Table \ref{table:B},
the bag constant is listed for different $\alpha$. We see that it only
weakly depends on $\alpha$.
It means that the bag constant is relatively stable for varying $\alpha$,
which is consistent with the ranges in Ref. \cite{PhysRevD.46.3211,LU1998443}.
\begin{table} [H]
	\caption{The bag constant}
	\begin{center}
		\begin{tabular}{c|c|c|c|c|c|c}
			\hline
             \hline
			 $\alpha$ & $0.5$  &  $0.6$  & $0.7$  & $0.8$  & $0.9$  &$1.0$\\\hline
			 $B^{1/4}$ $(\mathrm{MeV})$       &132.918     & 133.37     &  134.672   & 136.94    & 141.09    &   151.91\\\hline \hline
		\end{tabular}
	\end{center}
\label{table:B}
\end{table}

By considering the thermodynamic relation, we have  \cite{PhysRevD.51.1989,PhysRevD.86.114028}:
\begin{align}
\varepsilon=-P+\rho_u\mu_u+\rho_d\mu_d.   \label{eq:22}
\end{align}
The total pressure and energy density includes the contributions from electrons:
\begin{align}
&P_{tot}=P_{quark}+\dfrac{\mu_e^4}{12\pi^2},\nonumber\\
&\varepsilon_{tot}=\varepsilon_{quark}+\dfrac{\mu_e^4}{4\pi^2}.\label{eq:23}
\end{align}
\begin{figure}[H]
	\centering
	\includegraphics[width=1\linewidth]{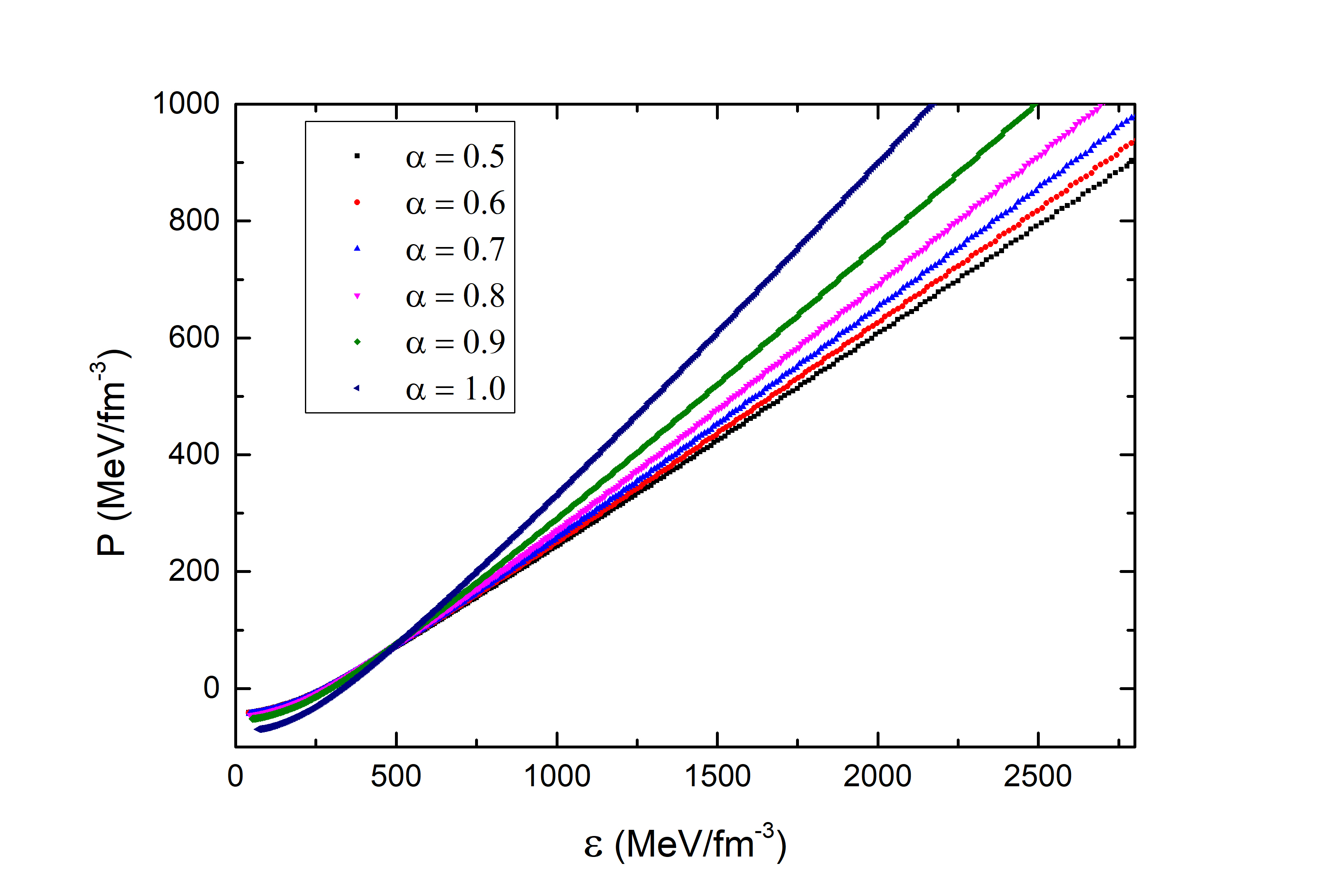}\\
	\caption{The equation of state.}\label{fig:EOS}
\end{figure}

\begin{figure}[H]
	\centering
	\includegraphics[width=1\linewidth]{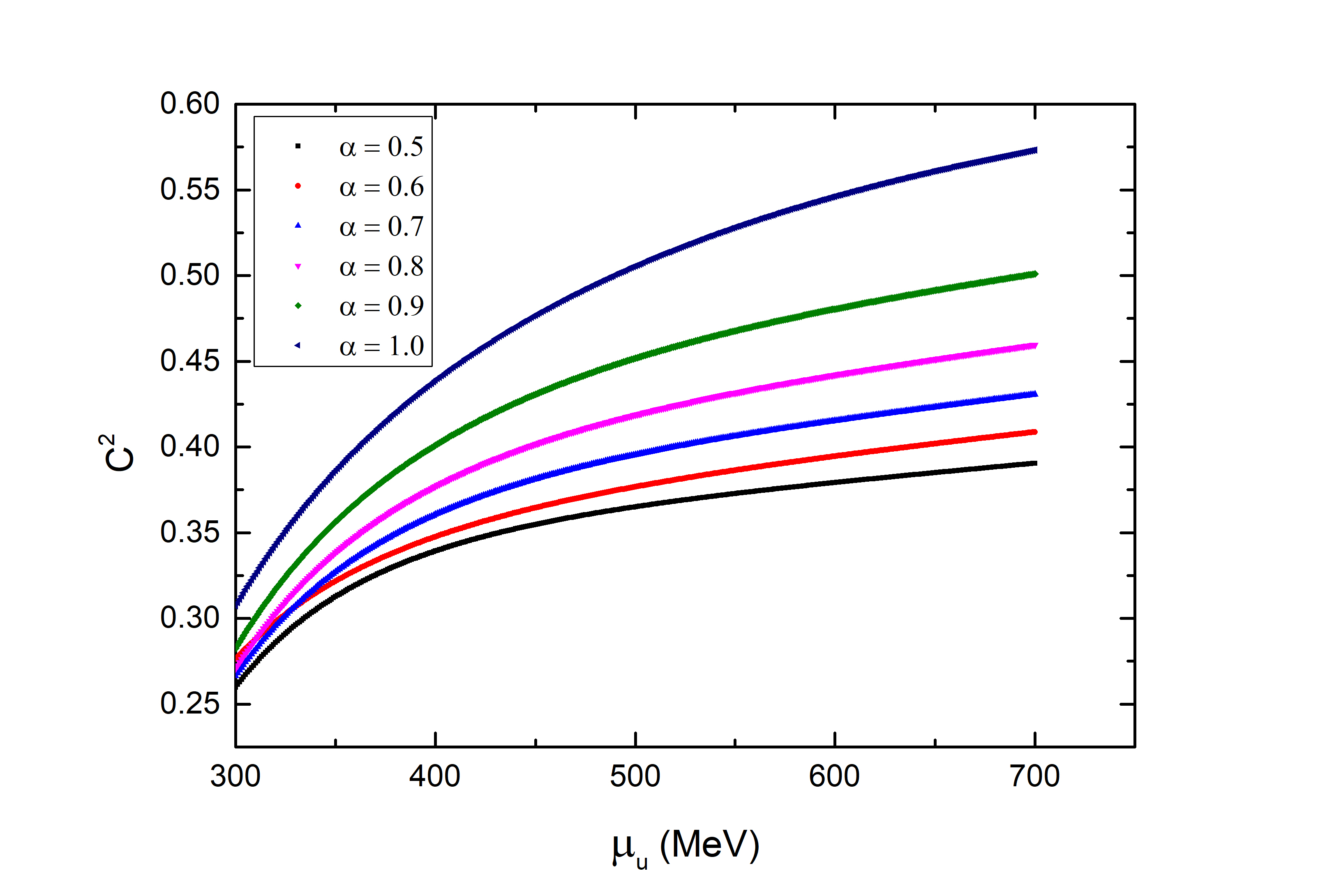}\\
	\caption{The square of the sound velocity as a function of the chemical potential.}\label{fig:sound}
\end{figure}

The relation between pressure and energy density (EOS) is illustrated in Fig. \ref{fig:EOS}. It shows that
the parameter $\alpha$ associated with the vector interactions makes the quark matter able to support larger pressures.
Besides, the stiffness of the quark star is indicated by the slope of the EOS curve. In Fig. \ref{fig:sound}, we
plot the square of the sound velocity, i.e.
\begin{align}
C^2=\dfrac{dP}{d\varepsilon}.\label{eq:24}
\end{align}
As the parameter $\alpha$ increases, the square of the sound velocity rises as well. In other words, the
vector interaction channel in the NJL model makes the EOS stiffer.

\section{THE TOV EQUATION}\label{section:5}

\begin{figure}[H]
	\centering
	\includegraphics[width=1\linewidth]{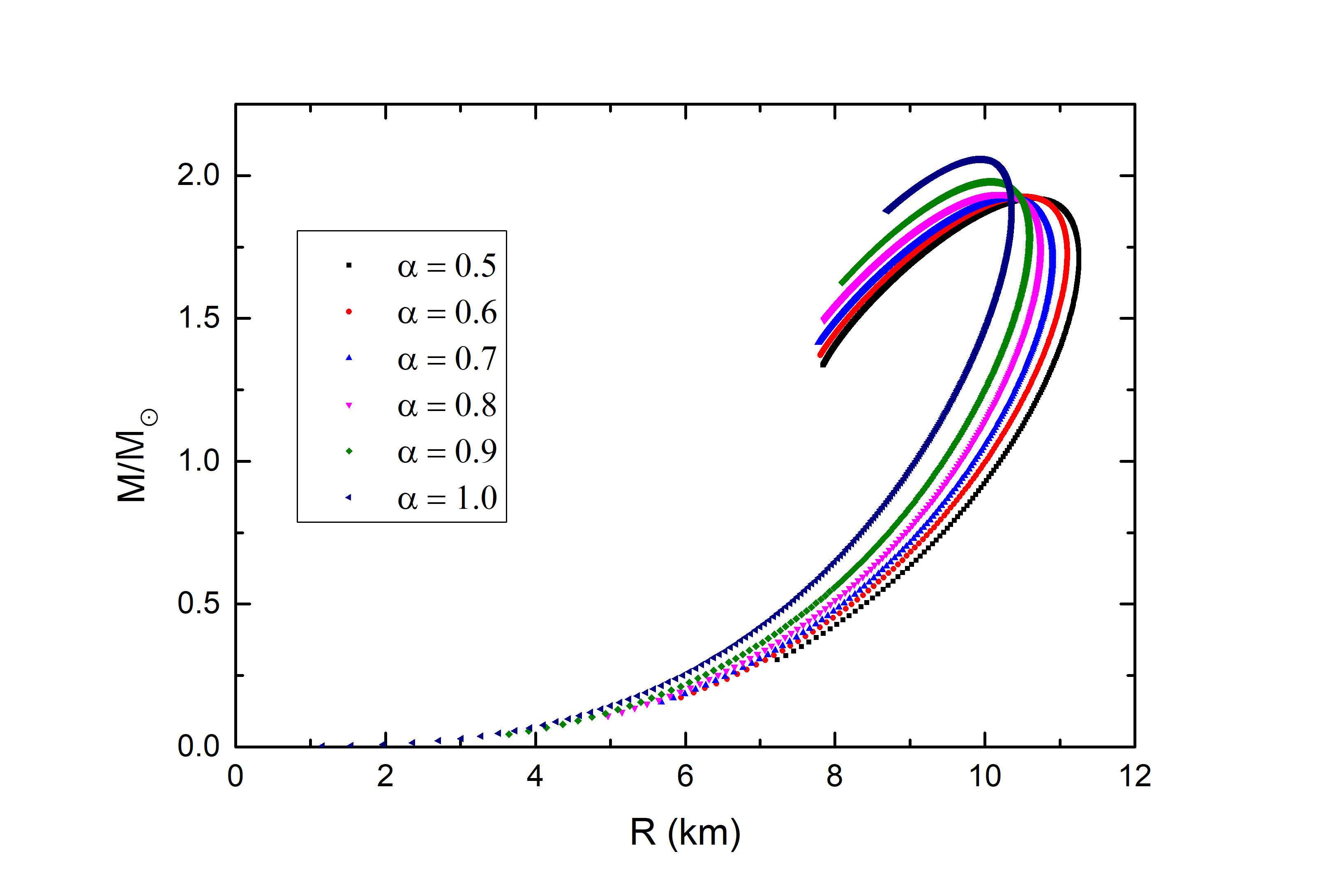}\\
	\caption{The mass-radius relation of quark stars.}\label{fig:massradius}
\end{figure}

Now that the EOS has been achieved, the Tolman-Oppenheimer-Volkoff equations (TOV) can be adopted to
calculate the structure of the compact star (as G = c = 1):
\begin{align}
&\dfrac{dP}{dr}=-\dfrac{(\varepsilon+P)(M+4\pi r^3 P)}{r(r-2M)},\label{eq:25}\\
&\dfrac{dM}{dr}=4\pi r^2\varepsilon.\label{eq:26}
\end{align}
The mass-radius relation is obtained by solving these differential equations. The results are
illustrated in Fig. \ref{fig:massradius}. We see that the parameter $\alpha$ advances the maximum mass of the quark star
but decreases the radius. When $\alpha$ equals to one, the maximum mass reaches 2.06-solar mass
with a radius of 9.9 km (Table \ref{table:mass}). It agrees with the observational constraints on the radius of less
than 13.6 and 13.76 km \cite{PhysRevLett.120.172702,PhysRevLett.120.172703}.

\begin{table} [H]
	\caption{The mass-radius relation for the maximum mass star}
	\begin{center}
\setlength{\tabcolsep}{7mm}{
		\begin{tabular}{c|ccc}
			\hline
\hline
			 $~\alpha~$ & $~M_{max}/M_{\odot}~$  &  $~R(km)~$    \\\hline
			 ~0.5~&~1.91~ &~ 10.7 ~\\
             ~0.6~&~1.92~ & ~10.5 ~\\
             ~0.7~&~1.92~ & ~10.3~ \\
			 ~0.8~&~1.94~ & ~10.2~  \\
			 ~0.9~&~1.98~ & ~10.1~\\
			 ~1.0~&~2.06~ &~9.9~ \\\hline \hline
		\end{tabular}  }
	\end{center}
\label{table:mass}
\end{table}

\section{THE TIDAL deformability}\label{section:6}

The tidal deformability is derived by the tidal Love number $k_2$,
and the Love number measures the distortion of the shape of the surface of a star by an external gravity.
In the unit $G=c=1$, the relation between $k_2$ and the tidal deformability is:
\begin{align}
k_2=\dfrac{3}{2}\Lambda(\dfrac{M}{R})^5.\label{eq:27}
\end{align}

The $l=2$ tidal Love number $k_2$ is calculated by \cite{Hinderer:2009ca}:
\begin{align}
k_2=&\dfrac{8}{5}C'^5(1-2C')^2\big[2+2C'(y-1)-y\big]\nonumber\\
&\times \big\{2C'[6-3y+3C'(5y-8)]\nonumber\\
&+4C'^3[13-11y+C'(3y-2)+2C'^2(1+y)]\nonumber\\
&+3(1-2C')^2[2-y+2C'(y-1)]ln(1-2C')\big\}^{-1},\label{eq:28}
\end{align}
where $C'=M/R$ is the compactness of the quark star and
\begin{align}
y=\dfrac{R\beta(R)}{H(R)}-\dfrac{4\pi R^3 \varepsilon_0}{M},\label{eq:29}
\end{align}
where $\varepsilon_0$ represents the surface energy density of the quark star.
The dimentionless value $y$ is obtained by two differential equations
\begin{align}
&~~~~~~~~~~~~~~~\dfrac{dH}{dr}=\beta\label{eq:30}, \\
\dfrac{d\beta}{dr}=&\dfrac{2H}{1-2M/r}\big\{-2\pi\big[5\varepsilon+9P + d\varepsilon/dP(\varepsilon+P)\big]\nonumber\\
&+\dfrac{3}{r^2}+\dfrac{2}{1-2M/r}(\dfrac{M}{r^2}+4\pi rP)^2\big\}\nonumber\\
&+\dfrac{2\beta}{r-2M}[-1+\dfrac{M}{r}+2\pi r^2(\varepsilon-P)].\label{eq:31}
\end{align}
As $r$ $\rightarrow  0$, $H(r)=a_0 r^2$ and $\beta(r)=2a_0 r$. $a_0$ can be any number,
because we only concern about the ratio between H and $\beta$.
In Table \ref{table:lambda}, the compactness, $k_2$
and the tidal deformability $\Lambda$ for the 1.4-solar-mass quark star are listed for different $\alpha$.

\begin{table} [H]
	\caption{Tidal deformability for 1.4-solar-mass quark star}
	\begin{center}
\setlength{\tabcolsep}{6mm}{
		\begin{tabular}{c|ccc}
			\hline
\hline
			 $~\alpha~$ & $~M/R~$  &  $~k_2~$  & $~\Lambda~$  \\\hline
			 ~0.5~&~0.1875~ &~ 0.1675 ~& ~481.812 ~\\
             ~0.6~&~0.1903~ & ~0.1667 ~& ~445.75 ~\\
             ~0.7~&~0.1936~ & ~0.1603~ & ~392.556 ~\\
			 ~0.8~&~0.1974~ & ~0.1561~ & ~347.466~ \\
			 ~0.9~&~0.2010~ & ~0.1542~ & ~313.1 ~\\
			 ~1.0~&~0.2083~ &~0.1487~ &  ~252.886~\\\hline \hline
		\end{tabular}  }
	\end{center}
\label{table:lambda}
\end{table}

The binary neutron star merger event GW170817 \cite{PhysRevLett.119.161101} gives us constraints on $\Lambda$.
For a 1.4 $M_{\odot}$ star, the tidal deformability should be  $\Lambda<800$ \cite{doi:10.1063/1.5117803}.
Table \ref{table:lambda} shows that this requirement is satisfied by our quark star model.

\section{SUMMARY AND CONCLUSIONS}\label{section:7}

In this paper, the EOS of deconfined quark stars is studied in the framework of the NJL model.
The mass-radius relation and the tidal deformability are obtained.
The self-consistent NJL model is employed by introducing the weighting factor $\alpha$ of
the Fierz-transformed Lagrangian. To obtain a stiffer quark star, a larger $\alpha$ will be needed.
Additionally, the proper-time regularization is used due to the large density in the core of quark stars, where the three-momentum cutoff assumption fails.
Because the past definition of the bag constant $B$ is not always suitable for the EOS of quark stars, for an analogy to the MIT bag model, we set the bag constant as the pressure of the quark matter that is undergoing the chiral phase transition along with the phase transition of the deconfinement.

The bag constant $B$ is found to vary from $(132.918 ~\mathrm{MeV})^4$ to $( 151.91 ~\mathrm{MeV})^4$ when $\alpha$ shifts from 0.5 to 1.0 under electric charge neutrality and $\beta$ equilibrium.  By solving the TOV equations,  the mass-radius relations for different $\alpha$ are yielded.
It is shown that as $\alpha$ increases, the maximum mass of quark stars grows but the corresponding radius decreases.
In particular, the maximum mass for $\alpha=1.0$ reaches 2.06-solar mass, which meets the observational requirement of
PSR J1614-2230 ($M=1.98\pm0.017 M_\odot$) and PSR J0348 + 0432 ($M=2.01\pm0.04 M_\odot$).
For a 1.4-solar-mass quark star, the derived radius is around 10 km, which is smaller than 13.6 km and 13.67 km
as required by observations \cite{PhysRevLett.120.172702,PhysRevLett.120.172703}.
Besides, the tidal deformability of quark stars in our framework ranges from 252.886 to 481.812, which also matches the
observational constraints of $\Lambda<800$ for a 1.4 $M_\odot$ compact star \cite{doi:10.1063/1.5117803}. Our study indicates that
deconfined quark stars may exist in the universe.

\section*{Acknowledgements}
We thank the anonymous referee for helpful comments and suggestions.
This work is supported in part by the National Natural Science Foundation of China 
(under Grants No. 12075117, No. 11535005, No. 11905104, No. 12005192, No. 11873030, 
No. 12041306, No. U1938201, and No. 11690030), by National SKA Program of 
China No. 2020SKA0120300, by China Postdoctoral Science Foundation (Grant 
No. 2020M672255 and No. 2020TQ0287), by the Strategic Priority Research Program of 
the Chinese Academy of Sciences (``multi-waveband Gravitational-Wave Universe'', 
Grant No. XDB23040000), and by Nation Major State Basic Research and Development 
of China (2016YFE0129300). In the end, we would like to thank Chao Shi for the 
constructive comments and reading through the manuscript.
\bibliography{ref}
\end{document}